\documentclass[final,english]{bullsrsl}[2022/06/15]

\usepackage[latin1]{inputenc}
\usepackage[T1]{fontenc}

\usepackage{natbib} 
\usepackage{graphicx}

\begin{document}
\title{Star formation characteristics of galaxies hosting AGN}

\author[affil={1,2},corresponding]{Payel}{Nandi}
\author[affil={2}]{C. S.}{Stalin}
\affiliation[1]{Joint Astronomy Programme, Department of Physics, Indian Institute of Science, Bangalore 560 012, India}
\affiliation[2]{Indian Institute of Astrophysics, Koramangala, Bangalore 560 034, India}

\correspondance{payel.nandi@iiap.res.in}
\date{today}
\maketitle


%

\begin{abstract}
We present an analysis of the ultra-violet (UV) observations of two Seyfert 
type active galactic nuclei (AGN), namely NGC$~$4051 and NGC$~$4151. These 
observations aimed at studying the star formation in the hosts of these AGN were 
carried out with the ultra-violet imaging telescope on board AstroSat in far-UV. 
A total of
193 and 328 star-forming regions (SF) were identified using SExtractor in
 NGC$~$4051 and NGC$~$4151, respectively. Using aperture photometry
of the identified SF regions, we estimated the star formation rates
(SFRs).  We found NGC$~$4051 to have the lowest SFR with a median value
of 3.16 $\times$ 10$^{-5}$ M$_{\odot}$ yr$^{-1}$ while for 
NGC$~$4151, we found a median SFR of 0.012 M$_{\odot}$ yr$^{-1}$.
\end{abstract}


\section{Introduction}
All massive galaxies host supermassive black holes (SMBHs) at their 
centers \citep{2013ARA&A..51..511K}. These SMBHs 
power active galactic nuclei (AGN) via accretion and are  believed to play an 
important role in the co-evolution
of SMBHs and their host galaxies. This belief is based on the observed relation 
between SMBH mass (M$_{BH}$)
and bulge mass \citep{1998AJ....115.2285M}, M$_{BH}$ and bulge luminosity 
\citep{2003ApJ...589L..21M} and M$_{BH}$ and  stellar velocity dispersion
\citep{2000ApJ...539L..13G,2000ApJ...539L...9F}. A mechanism that is invoked to understand
the co-evolutionary scenario in AGN is by feedback processes \citep{2012ARA&A..50..455F}. 
Therefore, finding observational signatures of AGN 
feedback is very important in enhancing our understanding of the effect of AGN on 
the evolution of their host galaxies. This triggers studies of star formation 
rate (SFR) in galaxies hosting AGN.  This is because, the connection between AGN and star formation (SF) activity in their hosts arises naturally as both the processes are fed by the same gas reservoir.

\begin{figure*}[t]
\centering
\hspace{0.3cm}\includegraphics[scale=0.65]{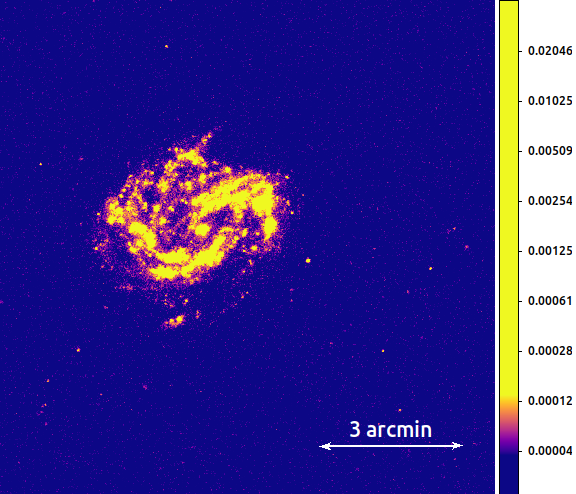}
\caption{The FUV image of NGC$~$4051. The brightness in count/sec is shown 
in the bar.}
\label{figure-1}
\end{figure*}

\begin{table*}
\caption{Details of the galaxies studied here. The $\alpha$, $\delta$, $z$ and 
morphology are from NED. The M$_{BH}$ for  NGC$~$4051 and NGC$~$4151 are 
from \cite{2009ApJ...702.1353D} and \cite{2022ApJ...934..168B}.} 
\label{table-1}
\bigskip
\begin{tabular}{cccccc} \hline
\textbf{Name}  &  \textbf{$\alpha$(2000)} & \textbf{$\delta$(2000)}  & \textbf{z}        &  \textbf{Morphology} &  \textbf{M$_{BH}$ (M$_{\odot}$)} \\ \hline
NGC$~$4051 &  12:03:09.64 &  +44:31:52.80    & 0.00234    &  SAB(rs)bc  & 1.73$^{+0.55}_{-0.52}$  $\times$ 10$^6$        \\
NGC$~$4151 &  12:10:32.58 & + 39:24:20.63    & 0.00333    &  SAB(rs)ab  & 1.66$^{+0.48}_{-0.34}$ $\times$ 10$^7$        \\ \hline
\end{tabular}
\end{table*}

\begin{figure*}[t]
\centering
\hspace{0.3cm}\includegraphics[scale=0.5]{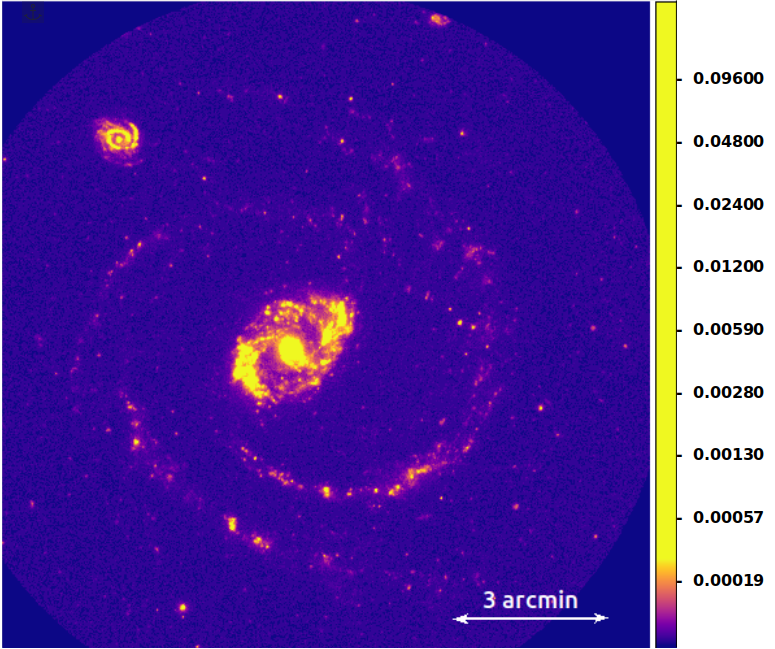}
\caption{The FUV image of NGC$~$4151. The brightness in count/sec is shown 
in the bar.}
\label{figure-2}
\end{figure*}

It is still poorly understood how AGN activity impacts star formation (SF) in their
host galaxies. However, there are evidences of both positive and negative feedback
in AGN. Negative feedback is when AGN emission is found to suppress SF and
is known in few systems \citep{2012MNRAS.425L..66M}. Alternatively, in sources 
with radio jets, there are evidences of the jets inducing SF, a case 
of positive feedback \citep{2020A&A...639L..13N,2017Natur.544..202M}. 
Also, studies of a few systems have revealed the co-existence of both positive and 
negative feedback on SF \citep{2019ApJ...881..147S,2021A&A...645A..21G} 
pointing to the complicated nature of AGN feedback.

Studies
available in the literature have addressed the issue of SF
around AGN (\citealt{2007ApJ...671.1388D} and references therein). Of those, 
ones that investigated the  SF characteristics 
close to the central SMBH on scales of a few hundred parsecs have 
focussed on Seyfert galaxies as they are the closest to us. Moreover,
one needs to probe different spatial scales to understand this connection better.
This is because, AGN and SF phenomena  occur on different spatial 
scales and time scales. The connection between AGN and SF can exist at 
smaller scales; however, at much larger scales, it is unlikely to find a connection
as those regions are too far to be influenced by the AGN. 
Therefore, to firmly understand the connection, if any, between AGN and SF, one 
needs to probe the SF characteristics of AGN host galaxies at 
different spatial scales using different probes.  We have been 
involved in a 
systematic investigation of a sample of Seyfert type 
AGN as part of an observational effort to characterize the SF in 
galaxies hosting AGN using data acquired in the ultra-violet band, with the Ultra-Violet Imaging Telescope (UVIT; 
\citealt{2017AJ....154..128T,2020AJ....159..158T}) 
onboard AstroSat, India's multi-wavelength astronomical observatory 
\citep{2006AdSpR..38.2989A}.  Here, we present our preliminary results on two 
sources based on UVIT observations.  Observations of these two sources in 
optical H$\alpha$ were recently acquired with the Devasthal optical 
telescope \citep{2019CSci..117..365S}, 
the results of which will be presented elsewhere.

\section{Observation and Data Reduction}
We observed NGC$~$4051 and NGC$~$4151 using 
UVIT. The details of the sources are given in Table \ref{table-1}.  
The observations were 
done in both far-UV (FUV; 1300 $-$ 1800 \AA) and near-UV
(NUV; 2000 $-$ 3000 \AA) bands. However, to characterize SFR, we used
observations acquired in FUV, the details of which are given in 
Table \ref{table-2}.  

We downloaded the Level 2 (L2) science ready images from the Indian Space 
Science Data Center (ISSDC). These images were processed by the 
UVIT-Payload Operations Center at the Indian Institute of Astrophysics
using the official L2 pipeline \citep{2022JApA...43...77G} 
and transferred
to ISSDC for archival and dissemination.
We redid astrometry on the L2 images using stars
available on the image frames with their ($\alpha$, $\delta$) values taken from
{\it Gaia-DR3} \citep{2023A&A...674A...1G}. These images, with new astrometry 
were used for further analysis. The final FUV image of NGC$~$4051 and NGC$~$4151 
is shown in Fig. \ref{figure-1} and Fig. \ref{figure-2}, respectively.

\section{Analysis and Results}
For all analysis in this work, we used the final image in F172M and F154M filters. To identify SF regions, we used the Source Extractor Software (SExtractor;  
\citealt{1996A&AS..117..393B}) by adopting the  parameters
of DETECT\_THRESH = 5 $\sigma$, DETECT\_MINAREA = 11 and DEBLEND\_THRESHOLD = 32.
We identified 193 SF regions in NGC$~$4051 and 328 SF regions in NGC$~$4151. 
We calculated the fluxes of the identified SF regions using the 
technique of aperture photometry for which we used PhotUtils 
\citep{2020zndo...4044744B}, after subtracting the background. 
To determine the background, we calculated the median count rate in square
boxes of size 50 $\times$ 50 pixels placed in 10 random regions devoid of
any UV sources. The median of these 10 measurements was taken as the 
background value per pixel which was then subtracted from each of the pixels
used to derive the fluxes of the SF regions. We note here that
while doing photometry of SF regions, we did not consider SF 
regions with overlapping apertures.

\begin{figure*}
\centering
\hbox{
    \includegraphics[scale=0.3]{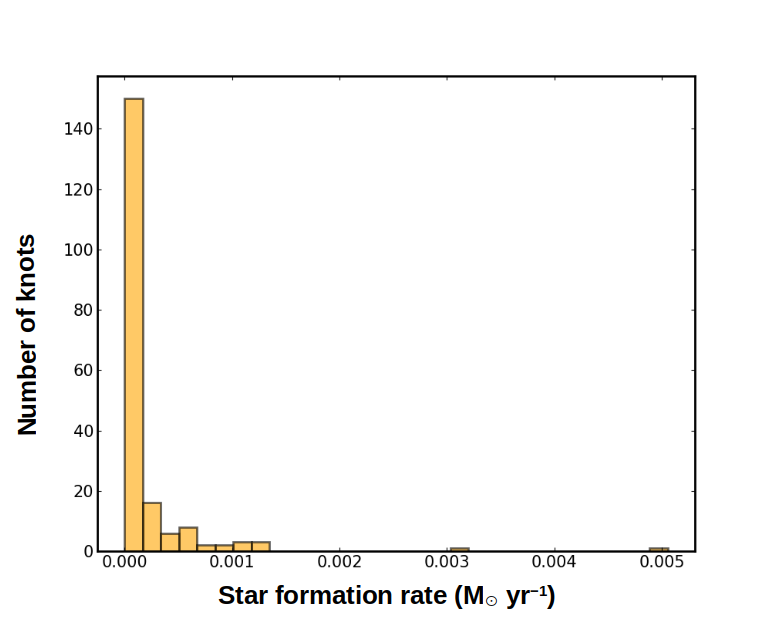}
    \includegraphics[scale=0.32]{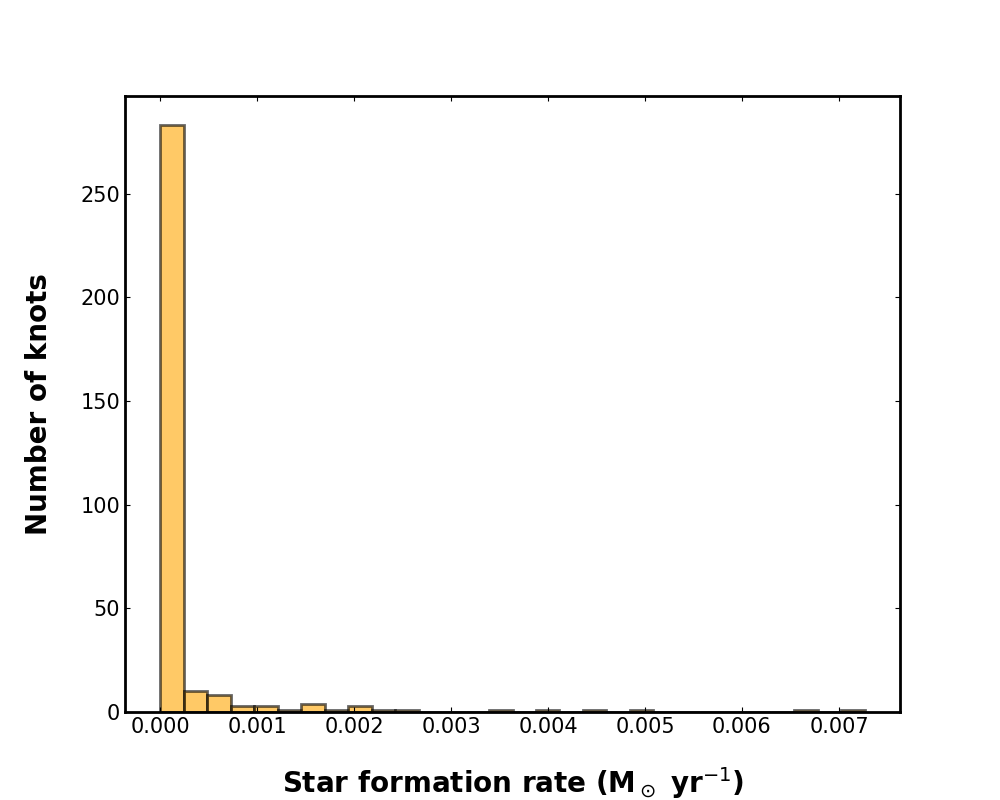}
    }
\caption{Distribution of SFR in NGC~4051 (left panel) and NGC~4151 (right panel).}
\label{figure-3}
\end{figure*}

We corrected the derived magnitudes of the SF regions for extinction
in UV following \cite{1989ApJ...345..245C}
\begin{equation}
<A(\lambda)/A(V)> = a(x) + b(x)/R(V)
\end{equation}
We adopted the Galactic extinction of A(V) = 0.036 mag and 0.074 mag in the V-band
for NGC$~$4051 and NGC$~$4151, respectively, from NED\footnote{https://ned.ipac.
caltech.edu}. In Equation 1, $x$ is the wave number and we calculated
a(x) and b(x) following \cite{1989ApJ...345..245C}. Similarly, to correct the
Galactic extinction corrected magnitudes for internal extinction, we used the $\beta$ 
method following the details outlined in \cite{2023ApJ...950...81N}. 
From the extinction corrected magnitudes, we calculated SFR as
follows \citep{2007ApJS..173..267S}.
\begin{equation}
log(SFR_{FUV}(M_\odot yr^{-1}))= log(L_{FUV}(W Hz^{-1})) -21.16
\label{eq:sfr}
\end{equation}
The distribution of SFR for NGC$~$4051 and NGC$~$4151 is  shown in 
Fig. \ref{figure-3}. Galaxies hosting AGN are known to have higher
SFR relative to those that do not host AGN. A possible interpretation of the
higher star formation activity in AGN hosts could be due to positive
feedback processes caused by AGN activity \citep{2012A&A...540A.109S}. 
However, the relation between
AGN and star formation is complex and various factors need to be considered
for a quantitative comparison of the SFR in galaxies with and without 
AGN \citep{2023A&A...675A.137M}.
The feedback process that might be at play in NGC 4051 and NGC 4151 is
being investigated (Nandi et al., in preparation).

\begin{table*}
\caption{Log of observations}
\label{table-2}
\bigskip
\begin{tabular}{cccc} \hline
\textbf{Name} & \textbf{Date of Observation} & \textbf{Filter} & \textbf{Exposure time}  \\ \hline
NGC$~$4051 &   08 June 2016, 11 Feb. 2018  &  F172M &    26444 sec   \\
NGC$~$4151 &   22 Feb. 2017, 17 Mar. 2017, 04 Jan. 2018, 02 May 2018  &  F154W &    67547 sec   \\ \hline
\end{tabular}
\end{table*}

\section{Conclusion}
In this work, we studied the SFR of two Seyfert galaxies, namely NGC~4051 and NGC~4151, the results of which are summarized below. 
\begin{enumerate}
\item We identified 193, and 328 SF regions in NGC~4051 and NGC~4151,  respectively.
\item For NGC~4051, we found the SFR to lie between 2.534$\times$ 10$^{-7}$ and 
50.55$\times$ 10$^{-4}$ M$_\odot$ yr$^{-1}$ with a median value of 
3.16$\times$ 10$^{-5}$ M$_\odot$yr$^{-1}$.
\item For NGC~4151, we found the SFR to lie between 4.35$\times$ 10$^{-4}$ to 
31.32$\times$ 10$^{-2}$ M$_\odot$ yr$^{-1}$ with a median value of 
1.188$\times$ 10$^{-2}$ M$_\odot$ yr$^{-1}$.
\end{enumerate}

\begin{acknowledgements}
The authors thank the anonymous referee for his/her
constructive comments on the manuscript, leading to its improvement. This publication uses the data from the AstroSat mission of the Indian Space Research Organization (ISRO), archived at the Indian Space Science Data 
Center (ISSDC). This publication uses UVIT data processed by the  payload 
operations center at IIA. The UVIT is built in collaboration between IIA, 
IUCAA, TIFR, ISRO, and CSA. PN thanks the Council of Scientific and Industrial Research (CSIR), Government of India, for supporting her research under the CSIR Junior/Senior research fellowship program through the grant no. $09/079(2867)/2021$-EMR-I.
\end{acknowledgements}

\begin{furtherinformation}

\begin{orcids}
\orcid{0009-0003-9765-3517}{Payel}{Nandi}
\orcid{0000-0002-4998-1861}{Chelliah Subramonian}{Stalin}
\end{orcids}

\begin{authorcontributions}
The first author reduced and analyzed the data. She wrote the first draft
of the manuscript. The second author contributed to the finalisation of the 
results and the manuscript.
\end{authorcontributions}

\begin{conflictsofinterest}
This manuscript is an outcome of an independent work. It does not conflict with 
others' ideas or results.
\end{conflictsofinterest}

\end{furtherinformation}

\bibliographystyle{bullsrsl-en}

\bibliography{ref}

\end{document}